\documentclass[twocolumn,showpacs,preprintnumbers,amsmath,amssymb]{revtex4}

\usepackage{graphicx}
\usepackage{dcolumn}
\usepackage{bm}

\begin{document}

\title{Study on Evolvement Complexity in an Artificial Stock Market}

\author{YANG Chun-Xia }
\author{ZHOU Tao}
\email{zhutou@ustc.edu}
\author{ZHOU Pei-Ling}
\author{LIU Jun }
\author{TANG Zi-Nan }2
\affiliation{ Department of Electronic Science and Technology,\\
University of Science and Technology of China, \\ Hefei Anhui,
230026, PR China }

\date{\today}

\begin{abstract}
An artificial stock market is established based on multi-agent .
Each agent has a limit memory of the history of stock price, and
will choose an action according to his memory and trading
strategy. The trading strategy of each agent evolves ceaselessly
as a result of self-teaching mechanism. Simulation results exhibit
that large events are frequent in the fluctuation of the stock
price generated by the present model when compared with a normal
process, and the price returns distribution is L\'{e}vy
distribution in the central part followed by an approximately
exponential truncation. In addition, by defining a variable to
gauge the ``evolvement complexity" of this system, we have found a
phase cross-over from simple-phase to complex-phase along with the
increase of the number of individuals, which may be a ubiquitous
phenomenon in multifarious real-life systems.
\end{abstract}


\maketitle

In social systems, such as insect societies, increased colony size
is associated with profound and wide-ranging changes in
``internal" organization and operation\cite{Bourke,Anderson}. For
instance, larger colony size is correlated with increased
homeostasis, cooperative activity, spatial organization of work,
and caste polymorphism to name but a few ``social correlates"
\cite{Bourke,Anderson}. Gautrais et al catch up with a model
demostrating a proximate mechanisms to emerge polymorphism in
insect societies, which indicates that specialization only occurs
above a critical colony size such that smaller colonies contain a
set of undifferentiated equally inactive individuals while larger
colonies contain both active specialists and inactive generalists,
as has been found in empirical studies\cite{Gautrais}. The
specialization of workers upon certain tasks can increase colony
productivity. The experimentation Weidenm\'{u}ller made indicated
that the dynamics of the colony response changed as colony size
increased: colonies responded faster
 to perturbations of their environment when they
were large (60 or more individuals) than when they were
small\cite{Weidenmller}. These findings provide intriguing new
examples of the ways in which individuals, each using only local
information, acting simply and independently and not subject to
any central or hierarchical control, can coordinate group-level
behavior which differs from that of each individual, as does
economical system. Every economical agent behaves simply contrast
to the system which is composed of them. Economical system complex
behaviors also result from repeated nonlinear interaction between
each others. But, does the social economical system have
similarity to insect societies that macro-properties have to do
with participants size? In this letter, we have found a phase
cross-over from simple-phase to complex-phase along with the
increase of the number of individuals based on our model, which
may be a ubiquitous phenomenon in multifarious real-life systems.

There are many modelling methods to explain origins of the
observed behavior of market price as emerging from simple
behavioral rules of a large number of heterogeneous market
participants, such as behavior-mind model\cite{Thaler,A},
dynamic-games model \cite{32}, multi-agent model \cite
{18,19,20,21,22,3,Wang1} and so on. The mainstream method is
agent-based modelling because of its simpleness, agility and
verisimilitude and which based on a stylized description for the
behavior of agents. Here, we proposed a stock market model based
on multi-agent that incorporates the feedback between the price
trend and agent's trading strategy. Therefore, our model will
demostrate that each agent has a limit memory of the history of
stock price and will choose an action according to his memory, and
that the trading strategy of each agent evolving ceaselessly as a
result of self-teaching mechanism will influence the price trend
inversely, which resemble the minority game\cite{4}.

In our model, before a trade, each agent should choose an action:
to buy, to sell or to ride the fence, the former two should
determine the price and amount of the trading-application. The
buyer with higher price and the seller with lower price will trade
preferentially, and the trading-price is the average of
selling-price and buying-price. The stock price is the weighted
average of trading-price according to the corresponding
trading-amount\cite{Xu}.

Each agent holds a so-called decision-matrix, which can tell him
how to do according to the history of stock price. Let $p(t)$ be
the stock price at time step $t$, then the range of fluctuation is
\begin{equation}
f(t)=(p(t)-p(t-1))/p(t-1)\in (-1, +\infty )
\end{equation}
For the sake of simplification, the range of fluctuation is
categorized into 5 types: drastic fall ($f\in (-1, -0.05)$), fall
($f\in [-0.05, -0.01]$), near immovability( $f\in (-0.01, 0.01)$),
rise ($f\in [0.01, 0.05]$) and drastic rise ($f\in (0.05, +\infty
)$), which are denoted by -2, -1, 0, 1 and 2 respectively. The
agent's memory is limited to the current 5 fluctuations, thus
there are $5^{5}=3125$ different fluctuation-patterns. The
decision-matrix contains the probabilities of trading strategies
according to the different fluctuation-patterns. For instance,
table 1 shows a decision-matrix of an agent named John. Based on
this matrix, John will choose to sell half of his shares in hand
at probability 0.40 when the present fluctuation-pattern is ``1,
0, -1, 0, 2". If an agent decides to buy or to sell, the
buying-price or selling-price will be chosen completely randomly
in the interval $[p(t), 1.1p(t)]$ or $[0.9p(t), p(t)]$
respectively.

\begin{table}
\caption{\label{tab:table1} John's decision matrix. A. sell all
his shares; B. sell half of his shares; C. do nothing; D. spend
all his cash on shares; E. spend half his cash on shares }
\begin{ruledtabular}
\begin{tabular}{@{}cccccc@{}}
Patterns &
Action A & Action B & Action C & Action D & Action E \\

$\cdots$ & $\cdots$ & $\cdots$ & $\cdots$ & $\cdots$ & $\cdots$\\
$1,0,-1,0,2$ & $0.10$ & $0.40$ & $0.20$ & $0.10$ & $0.20$\\
$\cdots$ & $\cdots$ & $\cdots$ & $\cdots$ & $\cdots$ & $\cdots$\\

\end{tabular}
\end{ruledtabular}
\end{table}

After a trade, each decision-matrix will change as a result of
self-teaching mechanism. For each agent, if his action made his
money increase, the corresponding probability in his
decision-matrix will be doubled, contrarily, it will be halved.
After that, the probabilities under the very pattern will be
normalized. For instance, if John's action were an unsuccessful
one, the probabilities under the fluctuation-pattern ``1, 0, -1,
0, 2" would become ``0.125, 0.25, 0.25, 0.125, 0.25" after
normalization. Apparently, there will be no changes if the agent
did nothing or his action kept his money unaltered. In order to
mimic the ``bounded rationality" and ``inductive thinking" of
investors\cite{Simon,Arthur2}, we set a very small probability
$\gamma$, which is called the reversal parameter. Agents may
change their decision-matrix in completely contrary direction at
the probability $\gamma$.

When proper initial condition and parameters have been chosen, the
artificial stock market can generate its stock price. In figure 1,
we report a typical simulation result about price time series
generated by our model, Which is similar to the reality (inset).
In this simulation we set the market size as $1000$ (i.e. 1000
stockholders), the initial stock price as $50$. The initial
quantity of fund and shares owned follows uniform distribution in
the interval $[0, 1000000]$ and $[0, 10000]$ respectively, and the
original fluctuation-pattern are randomly selected from the 3125
candidates. Notice that an agent's action may be restricted by his
wealth. In other words, he may be prevented from buying or selling
because of, respectively, a shortage of fond or shares in hand.

\begin{figure}
\scalebox{0.7}[0.8]{\includegraphics{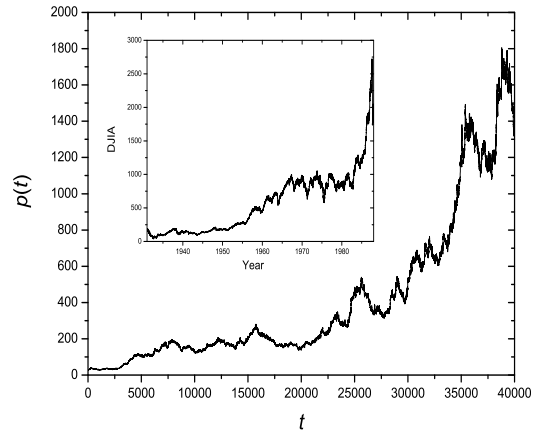}}
\caption{\label{fig:epsart} Time series of the typical evolution
of the stock price in the interval $t\in [0,40000]$, where $\gamma
=0.01$ and the elements of initial decision-matrices are chosen
completely randomly in the range $[0, 1]$ before normalization.
The insect is the Dow Jones Industrial Average(DJIA) from
01-02-1931 to 12-31-1987.}
\end{figure}

Large numbers of simulations have been performed to check if the
model can generate price time series of key characteristics
according with the reality. Since chaotic characteristic is one of
complex dynamical properties of economical system, which has been
demonstrated by previous studies\cite{Peters,Philipatos}. We have
calculated the Lyapunov exponent and correlative dimension of the
stock price time series generated by the model, carried out
principal component analysis, and drawn the conclusion that our
model can not only create stock price trends rather similar to the
real, but also show the chaotic behavior in deep consistency with
the real stock market. The details are omitted, and can be
referred to the corresponding reference\cite{Zhou}.

In addition, we have calculated the distribution of price return
$r(t)$, where $r(t)$ is defined as the difference between two
successive logarithms of the price:
\begin{equation}
r(t)=\beta(\texttt{log}p(t+\Delta t)-\texttt{log}p(t))
\end{equation}
Here, $\beta$ is a positive constant. The corresponding price
returns with $\Delta{t}=2$ are shown in figure 2, from which one
can see that large events are frequent in the fluctuation of the
stock price generated by the present model when compared with a
normal process, which agrees with the previous empirical studies
well\cite{Gopikrishnan,Mantegna,Galluccio,Liu,Wang2}. In figure 3,
we plot the probability distribution of price returns, and the
fitted Gaussian curve for the case $\Delta t=1$. Comparing with
normal distribution, the present returns distribution is of more
peaked center and fatter tail, according with the empirical
studies that suggest the distribution of returns in real-life
financial market is a L\'{e}vy distribution in the central part
followed by an approximately exponential
truncation\cite{Gopikrishnan,Mantegna,Galluccio,Liu,Wang2}. Since
our main goal in this letter is not to show the comparison between
price time series generated by our model and the reality, more
details are omitted here.

\begin{figure}
\scalebox{0.7}[0.8]{\includegraphics{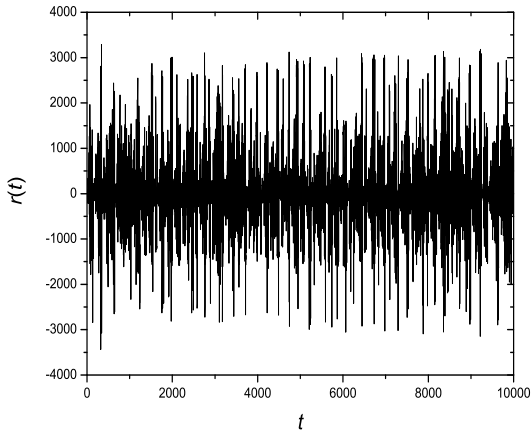}}
\caption{\label{fig:epsart} The price returns with $\Delta{t}=2$,
where $\beta=10^5$ and $0<t<10000$.}
\end{figure}

\begin{figure}
\scalebox{0.7}[0.8]{\includegraphics{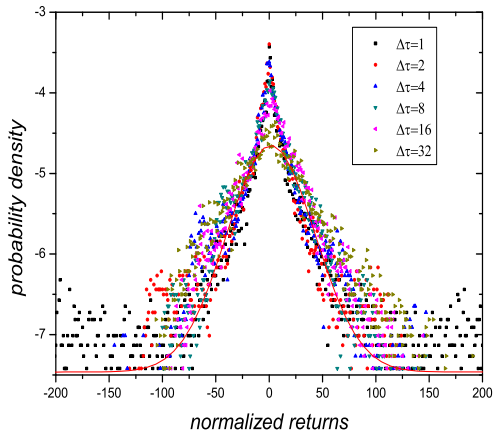}}
\caption{\label{fig:epsart} The normalized probability density of
price returns over different time scale $\Delta t=1,2,4,8,16,32$.
The solid curve is the fitted Guassean curve for the case $\Delta
t=1$.}
\end{figure}

In succession, let's discuss whether macro-properties have to do
with participants size. As far as whether the system is complex
and how much is the complexity degree are concerned, there isn't a
criterion. Whether the system is complex is sometimes apparent.
This, however, is not necessarily the case. If there is only one
simple time series generated by a certain system, it is very hard
to make sure that the system is complex. On the contrary, we can
easily find the evidence if the system is obvious incomplex. In
order to answer that problem , we loosely define a variable, i.e.
\emph{\textbf{evolvement complexity}} under the condition that the
behaviors of our artificial economical system are similar to those
of the reality. In general, our model will run long time. But the
model with small market size will run more shorter time than that
with large market size and will generate price time series which
tail is monotone or belongs to a finite period or reaches a
approximate fixed point. Such a price time series $p(t)$ generated
by our model with length $L$ is defined to be simple. Here,
``tail" means the last $0.1L$ points, and ``it reaches an
approximate fixed point" means the ratio of its range to its
average is smaller than 0.05. Let the market size be fixed and
other parameters be variable, if $m$ simple time series are
generated by $n$ independent experiments, then the evolvement
complexity of size $N$ is loosely defined as $C(N)=\frac{n-m}{n}$.
Note that the length of tail ($0.1L$) and the criterion of what is
an approximate fixed point ($<0.05$) are not especially selected
to generate the simulation results shown in the present letter.
One can write a program and easily check that the simulation
results are robustious for a wide parameter space, thus the
following phenomena are.

\begin{figure}
\scalebox{0.70}[0.8]{\includegraphics{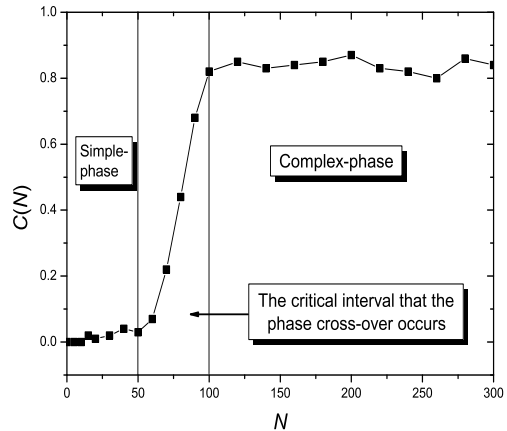}}
\caption{\label{fig:epsart}The evolvement complexity of size $N$,
where $L=10000$, $n=100$ and $\gamma$ is randomly chosen in the
interval $[0,0.1]$.}
\end{figure}

As is shown in figure 4, the system evolutive behavior is
evidently divided into 3 areas. When $N\leq 50$, the behavior is
simple, and its evolvement complexity increases rather slowly with
the increasing of the number of individuals. When $N\geq 100$, the
system has a great evolvement complexity, but its ``complex
degree" does not increase with the increasing of the number of
individuals. Therefore the areas $N\leq 50$ and $N\geq 100$ can be
considered as the simple-phase and complex-phase respectively.
Between these, the complexity increases fiercely with the
increasing of the number of individuals, thus it is called the
critical interval with the inf-critical-point 50 and
sup-critical-point 100. Here, we have found a phenomenon that the
behavior of our artificial economical system is correlated with
increased participants size, which is similar to that of insect
societies.

The economical system constitutes one among many other systems
exhibiting a complex organization and dynamics with similar
behavior, which, with large number of mutually interacting parts,
self-organize their dynamics with novel and sometimes surprising
macroscopic (``emergent") properties. The phenomenon that the
evolvement complexity can be divided into 3 parts owing to the
increasing of the number of individuals maybe one of the common
characteristics among various complex systems that do not seem
alike at all in appearance. The conception ``evolvement
complexity" is novel and interesting, but it is hard for us to
give out a strict and appropriate definition. Therefore, it is a
innovative point as well as a shortcoming in this letter. Although
the definition is rough, we believe it will enlighten physicists
on how to measure the complexity of complex systems.

This work is supported by the National Natural Science Foundation
of China under Grant No. 70171053, and the Foundation for graduate
students of University of Scicence and Technology of China under
Grant No. USTC-SS-0501.

\end{document}